\documentstyle[prd,aps,twocolumn,amsfonts,amssymb,amsmath]{revtex}
\newcommand{\aconn}{\mathcal{A}}

\newcommand{\caus}{\vec{P}}

\newcommand{\gauge}{\mathcal{U}}
\newcommand{\kd}{\text{\texttt{d}}}

\newcommand{\omg}{\Omega}
\newcommand{\qaus}{\vec{\omg}}

\newcommand{\cont}{{\mathcal{C}}^{0}}
\newcommand{\smooth}{\mathcal{C}^{\infty}}
\newcommand{\spn}{{\rm span}}

\newcommand{\com}{\mathbb{C}}
\newcommand{\A}{\mathbb{A}}
\newcommand{\mapto}{\rightarrow}
\newcommand{\maptoo}{\longrightarrow}

\newcommand{\N}{\mathbb{N}}
\newcommand{\Z}{\mathbb{Z}}
\newcommand{\R}{\mathbb{R}}

\begin{document}
\twocolumn[\hsize\textwidth\columnwidth\hsize\csname@twocolumnfalse\endcsname

\title{Quantum Space-Time as a Quantum Causal Set}

\author{Ioannis Raptis \vspace{3mm}}

\address{Theoretical Physics Group, Blackett Laboratory,
Imperial College of Science, Technology and Medicine,\\ Prince
Consort Road, South Kensington, London SW7 2BZ, UK \vspace{2mm}}

\maketitle

\begin{abstract}

A recently proposed algebraic representation of the causal set
model of the small-scale structure of space-time of Sorkin {\it et
al.} is briefly reviewed and expanded. The algebraic model suggested,
called {\em quantum causal set}, is physically interpreted as a
locally finite, causal and quantal version of the kinematical
structure of general relativity: the $4$-dimensional Lorentzian
space-time manifold and its continuous local orthochronous Lorentz
symmetries. We discuss various possible dynamical scenarios for
quantum causal sets mainly by using sheaf-theoretic ideas, and we
entertain the possibility of constructing an inherently finite and
genuinely $\smooth$-smooth space-time background free quantum
theory of gravity. At the end, based on the quantum causal set
paradigm, we anticipate and roughly sketch out a potential future
development of a noncommutative topology, sheaf and topos
theory suitable for quantum space-time structure and its dynamics.

\medskip



\end{abstract}

\vspace{8mm}]

The causal set approach to quantum gravity was initially proposed
by Sorkin and coworkers almost a decade and a half ago
\cite{ref1}. At the heart of this theoretical scenario lies the
proposal that the structure of space-time at quantum scales should
be modelled after a locally finite partially ordered set (poset)
of elements, the {\em causal set}, whose partial order is the
small-scale correspondent of the relation that defines past and
future distinctions between events in the space-time continuum of
macroscopic physics. It is remarkable indeed that from so simple
an assumption one can recover the basic kinematical features of
the classical space-time manifold of general relativity, namely,
its topological ({\it ie}, $\cont$-continuous), its differential
({\it ie}, $\smooth$-smooth) and its conformal Lorentzian metric
structures, as well as, in a statistical and scale dependent
sense, its dimensionality. That alone should suffice for taking
the causal set scenario seriously.

However, by abiding to a Wheeler-type of principle holding that
{\em no theory can qualify as a physical theory proper unless it
is a dynamical theory}, one could maintain that causal set theory
would certainly be able to qualify as a physically sound
theoretical scheme for quantum gravity if it somehow offered a
plausible dynamics for causal sets. Thus, `{\em how can one vary a
locally finite poset?}' has become the central question underlying
the quest for a dynamical theory of causal sets \cite{ref2}. To be
sure, a classical stochastic sequential growth dynamics for causal
sets, supported and guided by a discrete analogue of the principle
of general covariance of general relativity, has been proposed
recently \cite{ref3} and it has been regarded as the stepping
stone to a deeper quantum dynamics which, in turn, has already
been anticipated to involve an as yet unknown `sum-over-causal
set-histories' argument \cite{ref1,ref4}.

In this paper we briefly revisit a recently proposed algebraic
picture of the causal set model of the kinematical structure of
space-time in the quantum deep, which is coined {\em quantum
causal set} and bears a sound quantum physical interpretation.
Some possible dynamical scenarios for these finite dimensional
noncommutative algebraic structures are sketched mainly along
sheaf-theoretic lines. For expository fluency and continuity, we
have decided to present the basic tenets of quantum causal set
theory by means of a brief history of the central mathematical and
conceptual developments that led to the quantum causal set idea.
At the end we venture into some novel mathematical structures that
are motivated by the quantum causal set paradigm, which are
cumulatively referred to here as {\em `noncommutative
topologies'}, and we discuss how they may be applied to the
problem of the quantum structure and dynamics of space-time.

The topological ancestors of causal sets are the so-called {\em
finitary substitutes of continuous ({\it ie}, $\cont$) manifolds},
which are $T_{0}$-topological spaces associated with locally
finite open covers of a bounded region $X$ in a topological
manifold space-time $M$, and which have the structure of posets
\cite{ref5}. To recall briefly how such posets arise, for $X$ a
bounded region in the $\cont$-manifold $M$ and $\gauge$ a locally
finite open cover of it, define the following {\em preorder} ({\it
ie}, reflexive and transitive) binary relation `$\rightharpoonup$'
between its points

\begin{equation}\label{eq1}
x\rightharpoonup y\Leftrightarrow x\in\Lambda(y) \Leftrightarrow
\Lambda(x)\subseteq\Lambda(y)
\end{equation}

\noindent with $\Lambda(x)|_{\gauge}:=\bigcap\{ U\in{\gauge} :\,
x\in U\}$. The preorder `$\rightharpoonup$' becomes a {\em partial
order} `$\mapto$' in the quotient space $P:=X/\leftrightarrow$,
where `$\leftrightarrow$' is the following equivalence relation
between $X$'s points

\begin{equation}\label{eq2}
x\leftrightarrow y\Leftrightarrow (x\rightharpoonup y)\wedge
(y\rightharpoonup
x)\Leftrightarrow\Lambda(x)|_{\gauge}=\Lambda(y)|_{\gauge}
\end{equation}

\noindent The relation $x\mapto y$ in $P$ (which $P$ really
consists of `$\leftrightarrow$'-equivalence classes $[x]$ of
points in $X$) can be literally interpreted as {\em the
convergence of the constant sequence $(x)$ to $y$ in the
$T_{0}$-topology of $P$} \cite{ref5}.

Such posets can be also viewed as {\em simplicial complexes} in
the homological sense of \v{C}ech-Alexandrov as {\em nerves of
open coverings of manifolds} \cite{ref6}, and the substitutions of
$X$ by them are regarded as {\em cellular approximations of a
$\cont$-manifold}---they are discretizations of the locally
Euclidean topology of $X\subset M$. Indeed, that they qualify as
approximations proper of the topological space-time manifold rests
on the fact that an inverse system of those posets `converges', at
the projective limit of infinite localization of $X$'s points by
`infinitely small' open sets about them, to a space that is
homeomorphic to $X$ itself \cite{ref5}.

Subsequently, the aforesaid finitary substitutes were represented
by finite dimensional, complex, noncommutative, associative {\em
incidence Rota algebras} and the resulting algebraic structures
were interpreted quantum mechanically \cite{ref6}. The standard
representation of a general poset ${\mathcal{P}}=(S,\mapto)$ by an
incidence algebra $\omg$, where $S$ is a set of elements and
`$\mapto$' a reflexive, antisymmetric and transitive binary
relation ({\it ie}, a partial order) between them, is given by

\begin{equation}\label{eq3}
\begin{array}{c}
{\mathcal{P}}=\{ p\mapto q:~p,q\in S\}\maptoo\cr \omg({\mathcal
P}):= \spn_{\com}\{ p\mapto q\} , \cr (p\mapto q) \cdot (r\mapto
s)= \left\lbrace
\begin{array}{rcl} p\mapto s &,& \mbox{if } q=r\cr 0 &,&
\mbox{otherwise}
\end{array} \right.
\end{array}
\end{equation}

\noindent which depicts the defining $\com$-linear and
(associative) multiplication structure of $\omg({\mathcal P})$,
noting also that associativity is secured by the transitivity of
`$\mapto$'.

The interpretation in \cite{ref6} of the incidence algebras
$\omg(P)$ associated with the finitary $T_{0}$-posets $P$ above as
{\em discrete quantum topological spaces} rests essentially on the
following four structural issues:

\begin{itemize}

\item In the Rota algebraic environment, the partial order
arrow-connections `$\mapto$' between the elements of the posets,
that actually define the aforementioned $T_{0}$-topologies on
them, can {\em superpose coherently with each other}. This
possibility for coherent quantum interference of topological
connections, which is encoded in the $\com$-linear structure of
the incidence algebras, is manifestly absent from the
corresponding posets which are merely associative multiplication
structures ({\it ie}, arrow semigroups or monoids, or even small
poset categories).

\item The incidence algebras are noncommutative.

\item The points extracted from these non-abelian algebras are,
in a technical sense, {\em quantum} and so are the topological
spaces that they constitute \cite{ref6,ref7}. On the one hand, the
qualification of points as being quantum comes from their being
identified with the kernels of (equivalence classes of)
irreducible (finite dimensional Hilbert space) representations of
the noncommutative incidence algebras, which kernels are in turn
primitive ideals in these algebras. On the other hand, the (Rota)
topology defined on these points can be thought of as being
quantum, because in the very definition of its generating relation
the noncommutativity of the algebras' product structure is
crucially involved. To recall briefly the relevant concepts and
constructions which are quite standard in algebraic geometry, and
which in the finitary context are cumulatively referred to as {\em
Gelfand spatialization} \cite{ref8}, one first defines points in
the incidence algebras, as follows

\begin{equation}\label{eq4}
\begin{array}{c}
\{\,\mbox{points in $\omg$s}\,\} =\cr \{\,\mbox{kernels of irreps
of $\omg$s}\,\} =\cr \{\,\mbox{primitive ideals in $\omg$s}\,\}
\end{array}
\end{equation}

\noindent Then one defines the generating relation `$\rho$' for
the so-called {\em Rota topology} on the primitive spectrum
$Spec(\omg)$ of $\omg$ in terms of $\omg$'s noncommutative
product, as follows

\begin{equation}\label{eq5}
\begin{array}{c}
I_{x},I_{y}\in Spec(\omg):~~ I_{x}\rho I_{y}\Leftrightarrow\cr
I_{x}\cdot I_{y} (\not=I_{y}\cdot I_{x})\subsetneqq I_{x}\cap
I_{y}
\end{array}
\end{equation}

\noindent where $I_{x}$ and $I_{y}$ are (kernels of) irreducible
(finite dimensional Hilbert space matrix) representations $x$ and
$y$ of $\omg$. In (\ref{eq5}), $I_{x}\cdot I_{y}$ is the product
ideal, while $I_{x}\cap I_{y}$ is the intersection ideal. For the
incidence algebras $\omg(P)$ associated with Sorkin's finitary
$T_{0}$-topological posets $P$, one can identify the indices of
the ideals $I_{x}$ and $I_{y}$ in $Spec(\omg(P))$ with the
vertices (points) $x$ and $y$ of $P$ by defining

\begin{equation}\label{eq6}
I_{x}:=\spn_{\com}\{ x\mapto y:~ (x\mapto y)\not= (x\mapto x)\}
\end{equation}

\noindent and verify that they are indeed ideals in $\omg(P)$. It
can be shown that the Rota topology is the weakest one in which
$I_{x}\rho I_{y}$ implies the convergence $x\mapto y$ of the point
$x$ to the point $y$ in $P$. This means that {\em `$\rho$' is the
transitive reduction of the partial order `$\mapto$' defining the
$T_{0}$-topology on $P$}. `$\rho$' corresponds to the so-called
covering relations in $P$. For more about the Gelfand
spatialization procedure and the non-standard Rota topology, the
reader should refer to \cite{ref6,ref7}.

Quantum points and topological spaces in the sense above have been
studied from the more general and abstract perspective of
mathematical structures representing `quantal topological spaces'
known as {\em quantales}. The latter were originally invented in
order to formulate a noncommutative version of the usual
Gelfand-Naimark representation theorem for abelian
$C^{*}$-algebras $\mathfrak{A}$ involving their maximal spectra
$Max(\mathfrak{A})$ ({\it ie}, the set of maximal ideals of $\mathfrak{A}$)
which are general, pointless, classical ({\it ie}, commutative)
topological spaces commonly known as {\em locales} \cite{ref9}.

\item The incidence Rota algebras and their finite dimensional
Hilbert space matrix representations have been seen to stand for
sound finitary-algebraic models of {\em space-time foam}
\cite{ref7}. The latter refers to the conception of {\em the
topology of space-time as a quantum observable}: a quantally
fluctuating, dynamically variable and in principle measurable
structure.

\end{itemize}

At this point it must be stressed that the incidence algebras
associated with Sorkin's finitary topological posets were shown to
encode combinatorial information not only about the topology of
the space-time manifold, but also about its {\em differential
structure}. In fact, they were seen to be {\em $\Z_{+}$-graded
discrete differential manifolds} in the sense that every $\omg(P)$
splits into the following direct sum of linear subspaces

\begin{equation}\label{eq7}
\begin{array}{c}
\omg(P)=\omg^{0}\oplus\omg^{1}\oplus\omg^{2}\oplus\cdots\cr
\omg^{i}:=\spn_{\com}\{ p\stackrel{\mathrm min}{\maptoo} q:~{\mathrm{deg}}
(p\stackrel{\mathrm min}{\maptoo}q)=i\in\N\}
\end{array}
\end{equation}

\noindent where ${\mathrm{deg}}(p\stackrel{\mathrm min}{\maptoo}
q)$ is the length of the shortest path $p\stackrel{\mathrm
min}{\maptoo}q$ connecting $p$ with $q$ ($\forall p,q\in P$) in
the Hasse diagram of $P$ and it corresponds to the homological
degree when the $P$s are viewed as simplicial complexes (nerves)
as alluded to above \cite{ref6,ref7}. $\omg^{0}$ is an abelian
subalgebra of $\omg$ called {\em the commutative algebra $A$ of
coordinates in $\omg$}, while
${\mathcal{R}}:=\bigoplus_{i\geq1}\omg^{i}$ is called {\em the
$A$-module of discrete differential forms in $\omg$}. There is
also a nilpotent K\"ahler-Cartan differential operator $\kd$,
which is defined in terms of the homological boundary $\delta$ and
coboundary $\delta^{*}$ operators when the $P$s are viewed as
simplicial complexes, satisfying a graded Leibniz rule and
effecting linear maps of the following sort

\begin{equation}\label{eq8}
\kd:~\omg^{i}\maptoo\omg^{i+1}
\end{equation}

\noindent But for more details the reader may again have to refer
to \cite{ref6,ref7} and references therein.

{\it In toto}, the $\omg(P)$s have been physically interpreted as
the reticular substitutes of bounded regions $X$ of the
$\smooth$-smooth space-time manifold $M$ of macroscopic physics.
Moreover, it has been argued that, similarly to the case of the
finitary posets, an inverse system of incidence algebras yields at
the projective limit of infinite localization of $X$ about its
points the commutative algebra ${\A}={}^{\com}{\smooth}(X)$ of
smooth complex-valued coordinates labelling $X$'s point events (as
it were, $A$ `converges' to $\A$), as well as the $\Z_{+}$-graded
$\A$-bimodule ${}^{\com}\omg$ of smooth complex differential forms
cotangent to every point event in the differential manifold
space-time (as it were, $\mathcal{R}$ `converges' to the module of
complex smooth exterior forms over $\A$)
\cite{ref6,ref7,ref10,ref11}.

At the same time, the aforementioned recovery of the classical
smooth space-time manifold, in the ideal limit of infinite
localization, from the quantal incidence algebraic substrata was
interpreted physically as {\em Bohr's correspondence principle},
or equivalently, as a {\em classical limit}, and it represented
the emergence of the classical local differential space-time
macrostructure from the `decoherence' of ({\it ie}, the breaking
of coherent quantum superpositions in) an ensemble of these
characteristically {\em alocal quantum space-time structures}
\cite{ref6,ref7}. Hence, the finite dimensional incidence algebras
associated with Sorkin's finitary poset substitutes of the
space-time continuum encode information about both the topological
($\cont$) and the differential ($\smooth$) structure of the
manifold. Arguably, this is an algebraic representation of Sorkin
{\it et al.}'s thesis mentioned earlier in connection with causal
sets that {\em a partial order effectively determines both the
topological and the differential structure of the space-time
continuum}.

On the other hand, in quite a dramatic change of physical
interpretation and philosophy nicely recollected in \cite{ref12},
Sorkin stopped thinking of the locally finite posets above as
representing finitary discretizations or simplicial decompositions
(with a strong operational flavor) of the space-time continuum
and, as noted in the opening paragraph, he regarded them as causal
sets. Then, he and coworkers posited that {\em the deep structure
of space-time is, in reality, a causal set} \cite{ref1}. Thus,
partial orders stand now for causal relations between events in
the quantum deep and not for topological relations proper, as it
were, the original `spatial' conception of a poset gave way to a
more `temporal' one. Of course, the ideas to model causality after
a partial order and that {\em at both the classical and the
quantum level of description of relativistic space-time structure,
partial order as causality is a more physical notion than partial
order as topology} have a long and noble ancestry \cite{ref13}.
So, as a consequence of Sorkin's `semantic switch', {\em the
macroscopic space-time manifold should be thought of as a coarse
approximation of the fundamental causal set substrata} in contrast
to the finitary topological posets which, as it was noted earlier,
were regarded as being coarse approximations of the space-time
continuum. Also, the local finiteness of the open covers of the
bounded region $X$ in the continuous space-time manifold $M$
relative to which the topological posets were defined in
\cite{ref5} was replaced by the slightly different local
finiteness of the causal sets, although the underlying
mathematical structure, the poset, remained the same
\cite{ref12,ref7}.

In the sequel, after having replaced the finitary topological
posets $P$ by causal sets $\caus$ as the truly fundamental
structures underlying the classical space-time manifold of
macroscopic experience, it was a rather natural step to associate
with the latter incidence Rota algebras $\qaus$ and, like in the
case of the incidence algebras $\omg(P)$ representing discrete
quantum topological spaces \cite{ref6,ref7}, to interpret the
resulting structures as {\em quantum causal sets} \cite{ref14}.
Hence, in the new non-abelian algebraic realm of quantum causal
sets, the causal arrows `$\mapto$' defining their causal set
correspondents can superpose or interfere quantum mechanically
with each other. This lies at the heart of the conception of {\em
quantum causality} and qualifies the epithet `quantum' in front of
`causal sets' used here \cite{ref7,ref14}. Here too, quantum
causal sets, like their causal set counterparts, stand for locally
finite, causal and {\em quantal} models of the kinematics of the
classical space-time of general relativity.

Having in hand sound reticular and quantal models of the
kinematics of gravitational space-time, and by abiding to the
aforesaid Wheeler-type of criterion for a physical theory, the
next stage in the development of quantum causal sets was the
search for a plausible dynamics for them. The basic idea in
\cite{ref15} was that in order to construct a dynamical theory of
quantum causality, and since Zeeman had shown that causality as a
partial order determines the conformal metric structure of the
{\em flat} ({\it ie}, {\em non-dynamical}) Minkowski space of
special relativity and its global orthochronous Lorentz symmetries
up to conformal transformations \cite{ref13}, one should somehow
look for a scenario to {\em curve quantum causality}. Then, the
main intuition, principally motivated by the fact that the curved
space-time of general relativity is Minkowski space localized or
gauged, was that we could possibly arrive at dynamical variations
of quantum causal sets by localizing (or gauging) them and,
concomitantly, by mathematically implementing this localization
process sheaf-theoretically. In other words, a possible reply to
Sorkin's question `{\em how can one vary a poset?}' mentioned
earlier is `{\em by sheaf-theoretic means}' \cite{ref16}. To this
end, we first needed to adapt basic sheaf-theoretic notions and
constructions to a locally finite setting. With that need in mind,
{\em finitary space-time sheaves} of (algebras of) continuous
observables on Sorkin's finitary substitutes were first defined in
\cite{ref17} as {\it \'{e}tal\'{e}s} spaces that are {\em locally
homeomorphic} to those $T_{0}$-posets $P$---the latter serving as
the topological base spaces for the sheaves.

Shortly after, and in a way analogous to the aforesaid transition
from finitary substitutes to causal sets and then to quantum
causal sets, {\em finitary space-time sheaves of the incidence
algebras modelling quantum causal sets} were defined in
\cite{ref15} along the lines of Mallios' Abstract Differential
Geometry (ADG). The latter pertains to the geometry of vector,
(differential) module and algebra sheaves which generalizes the
usual differential calculus on $\smooth$-manifolds to an abstract
differential geometry on, in principle, {\em any (topological)
base space} \cite{ref18}. {\em Principal finitary space-time
sheaves of quantum causal sets} over the causal sets $\vec{P}$ of Sorkin
{\it et al.}, having for structure group reticular versions of the
continuous local orthochronous Lorentz group manifold
$L^{+}=SO(1,3)^{\uparrow}$ of general relativity, together with
non-flat finitary $\ell^{+}=so(1,3)^{\uparrow}\simeq
sl(2,\com)$-valued spin-Lorentzian connection $1$-forms $\aconn$,
were seen to be sound models of a locally finite, causal and
quantal version of the kinematics of classical Lorentzian gravity
in its gauge-theoretic formulation in terms of the spin-Lorentzian
connection $\aconn$ on a principal fiber bundle over the
$\smooth$-smooth space-time manifold $M$ having $L^{+}$ as its
structure group. Furthermore, finitary-algebraic versions, of
strong categorical ({\it ie}, functorial) character, of the
principles of locality, equivalence and general covariance were
formulated entirely in sheaf-theoretic terms mainly due to the
fact that the $\aconn$s were represented by suitable {\em finitary
space-time sheaf morphisms}.

The next step in the development of quantum causal set theory was
to carry the entire de Rham theory of differential forms on the
classical $\smooth$-smooth space-time manifold, virtually
unaltered, to the reticular realm of the curved finitary
space-time sheaves of quantum causal sets with the help of some
rather `universal' sheaf-theoretic techniques, mainly based on
sheaf cohomology, borrowed from ADG \cite{ref19}. One of the main
achievements of that transcription was the {\em
sheaf-cohomological classification of the non-flat reticular
spin-Lorentzian connections} $\aconn$ the quanta of which,
originally coined {\em causons} in \cite{ref15}, were taken to
represent {\em dynamical quanta of causality}---elementary
particle-like entities, anticipated to be closely related to
gravitons, that dynamically propagate in a discrete manner in the
`{\em curved quantum space-time vacuum}' represented by the curved
finitary space-time sheaves of quantum causal sets.

We also showed how basic differential geometric ideas and results
usually thought of as being vitally dependent on $\smooth$-smooth
manifolds for their realization, as for example the standard de
Rham cohomology, carry through unchanged to the finitary regime of
the curved finitary space-time sheaves of quantum causal sets. For
instance, we gave reticular versions of central
$\smooth$-theorems such as de Rham's, Weil's integrality, as well
as the Chern-Weil theorem. By this essentially complete
transcription of the basic $\smooth$-constructions, concepts and
results to the locally finite and quantal regime of the curved
finitary sheaves quantum causal sets, we highlighted that for the
formulation of fundamental differential geometric notions the
classical smooth background space-time continuum is of no
significantly contributing value. Quite on the contrary, we argued
that since the $\smooth$-smooth space-time manifold can be
regarded as the main culprit for the singularities that plague
general relativity and the weaker but still troublesome infinities
that assail the flat quantum field theories of matter, its
evasion---especially by the finitistic-algebraic means that we
employed---should be most welcome for the formulation of a
`calculationally' and, in a sense to be explained below,
`inherently' finite and $\smooth$-smooth space-time background
independent quantum theory of gravity.

However, it is fair to say that, the aforementioned kinematical
developments aside, an explicit dynamics for quantum causal sets
has not been proposed yet. The formulation of a finitary, causal
and quantal analogue of the classical vacuum Einstein equations
for gravity, which in turn can be interpreted as the dynamical
equations for the reticular sort of Ashtekar's (self-dual)
spin-Lorentzian connection variable $\aconn$ inhabiting the
relevant curved principal finitary space-time (line) sheaves of
quantum causal sets, is currently under way again along the lines
of ADG \cite{ref20}. A `covariant' path integral over the space
$[{\aconn}]$ of connections $\aconn$, or an even more finitistic,
as befits our discrete models, `sum-over-quantum causal
set-histories' scenario appears to be {\it prima facie} more well
suited for such a quantum dynamics than a canonical ({\it ie},
Hamiltonian) approach mainly due to the characteristic absence in
the innately granular realm of the causal or the quantum causal
sets of (a generator of) infinitesimal time increments
\cite{ref2}. This also seems to accord with Sorkin {\it et al.}'s
anticipation noted earlier of a `sum-over-causal set-histories'
dynamical scheme for causal sets. Moreover, we anticipate that
precisely because of the discrete character of the spin-Lorentzian
connections dwelling on the principal finitary space-time sheaves
of the finite dimensional quantum causal sets, the problem of
finding a well-defined measure for the aforesaid path integral
over the infinite dimensional coset space (orbifold or `moduli
space') $[{\aconn}]/{\mathrm{Diff}}(M)$ of
diffeomorphism-equivalent connections defined on the
$\smooth$-smooth space-time continuum, may simply disappear in our
finitary model.

Another possible quantum dynamics for quantum causal sets, now
perhaps one with a more canonical flavor, arises when, instead of
working with the individual incidence algebras representing
quantum causal sets {\it per se}, one works with the `universal'
incidence algebra of all finitary incidence algebras that are
partially ordered by inclusion \cite{ref2}---again itself a poset
representing `quantum causal refinement' \cite{ref15}. This poset
may be thought of as the incidence algebraic analogue of the
lattice of all (finite) topologies studied in \cite{ref21} in the
context of {\em quantum topology}, and it may be interpreted as
some kind of kinematical state space for quantum causal sets. One
could then try to define suitable generalized position and
momentum-like {\em observables of quantum causal topology} {\it
\`a la} Isham, impose some non-trivial canonical commutation
relations on them, and look for a Hamiltonian-like operator in
terms of them that effects contiguous dynamical transitions
between quantum causal sets, as it were, {\em dynamical quantum
(causal) topology changes} in the universal incidence algebra.
This could also be done `by hand', that is, one could first define
creators and annihilators of vertices in causal sets, look for
their incidence algebraic correspondents in the associated quantum
causal sets and their finite dimensional Hilbert space
representations as in \cite{ref7}, then consider their `canonical'
commutation relations, and finally proceed in a Hamiltonian
fashion to implement dynamical quantum causal topology changes as
briefly alluded to above \cite{ref22}.

In parallel with these canonical scenarios, one could also try to
define up-front the notion of {\em quantum causal histories}
\cite{ref23} and the finitary sheaves thereof in the universal
incidence algebra \cite{rap2} and proceed in a
consistent-histories-like fashion to look for a {\em quantum
measure} for quantum causal histories, relative to which the
`sum-over-quantum causal set-histories' is going to be weighed, in
the form of some kind of {\em decoherence functional} on them
\cite{ref2}. Alternatively, one could seek directly for a {\em
discrete quantum causal dynamics} and for functorial expressions
of {\em discrete causal covariance} and {\em quantum entanglement}
by applying ideas from linear logic and the theory of
polycategories to quantum causal histories \cite{ref24}.

Certainly though, and in spite of the prominent absence of an
explicit quantum dynamics, whether covariant, canonical or other,
for either the causal or the quantum causal sets, the
finitary-algebraic and sheaf-theoretic approach to quantum
space-time structure supporting quantum causal set theory has
provided us with strong clues on how to construct {\em an
intrinsically finite quantum gravity that is genuinely
$\smooth$-smooth space-time background independent}. For one
thing, it has shown us that one can define the usual differential
geometric objects and carry out the standard smooth constructions
that are of vital importance for the formulation of general
relativity, such as {\em connection}, {\em curvature} and {\em de
Rham cohomology} to name a few, {\em literally without the use of
any Calculus} and {\em over reticular base spaces that may appear
to be unmanageably singular and incurably pathological from the
perspective of the classical $\smooth$-smooth space-time
manifold}. This is one of the benefits we get from applying ADG to
the finitary-algebraic setting, for ADG has time and again proven
itself when it comes to evading $\smooth$-smoothness and the
singularities that go with it \cite{ref19,ref20,ref25}. If
anything, such an independence of quantum causal set theory (and
of a future dynamics for it) from the smooth space-time continuum
is more than welcome from the point of view of modern research in
(canonical) quantum gravity, since the theory appears to avoid
{\it ab initio} the infinite dimensional diffeomorphism group
${\mathrm{Diff}}(M)$ of general relativity, which comes hand in
hand with the $\smooth$-smoothness of the space-time manifold $M$,
and the two notorious problems associated with it, namely, the
`{\em inner product problem}' and the `{\em problem of time}'
\cite{ref26}, as well as the path integral measure problem in the
${\mathrm{Diff}}(M)$-covariant approach mentioned earlier.

We conclude the present review by elaborating on how the quantum
causal set model and its sheaf-theoretic representations may
substantially motivate the development of a {\em noncommutative}
topology and an associated sheaf as well as topos theory for
quantum space-time structure and its dynamics \cite{ref27}. We
first note that it seems theoretically short-sighted and lame to
think that only a high-level structure such as the geometry of
space-time should be subject to dynamics and quantization, while
that a deeper structure like its topology should be fixed once and
forever by the theoretician to the immutable classical continuous
manifold. As noted earlier, the incidence Rota algebras which have
been employed to model quantum causal sets have also been used to
represent {\em space-time foam}---Wheeler's original insight that
not only the geometry but also the topology of space-time should
partake into dynamical changes and coherent quantum superpositions
\cite{ref7}.

It must be also emphasized that it appears unreasonably limited to
possess a well developed noncommutative differential geometry
\cite{ref28} enjoying numerous applications to quantum space-time
and gravity, while the topology, which again is a structure deeper
than the differential, to be treated essentially commutatively
(classically) \cite{ref29}. Another possible way to arrive at a
suitable noncommutative topology for quantum space-time and its
dynamics, now by using the sheaf-theoretic localizations of
quantum causal sets as in \cite{ref15}, could be the following:
Finkelstein and coworkers, as part of an ongoing effort to find a
quantum replacement for the space-time manifold of macroscopic
physics, have developed a theory of quantum sets, which in a sense
represents a quantization of ordinary `classical' set theory
\cite{ref30}. The basic idea is that space-time at small scales
should really be viewed as a `quantum' set, not a classical one.
This is supposed to be a step on the path to a `correct' version
of quantum space-time topology and quantum gravity. Now, a
question which may occur to a modern logician (or topos theorist)
is the following:

\begin{quotation}
\noindent {\em what is so special about the category ${\mathbf
Set}$ of classical sets, since there are other logical universes
just as good, and possibly better, namely `topoi'} \cite{ref31}?
\end{quotation}

\noindent Perhaps it would be a better idea to try and quantize
these more general categories, since the use of ${\mathbf Set}$
may be prey to classical chauvinism. That is, the underlying
logical model may turn out to be some kind of `{\em quantum
topos}' \cite{ref32} different from the category of sets whose
existence is after all predicated upon the preconceptions of
macroscopic thinkers.

For, arguably, all the usual flat classical and quantum field
theories of matter are conveniently formulated in ${\mathbf Set}$,
or more precisely, in ${\mathbf Shv}(M)$---the `classical' topos
of sheaves of sets over the classical space-time manifold $M$
\cite{ref31,ref32}. However, as noted earlier, these continuum
theories suffer from non-renormalizable infinities coming from
singularities that plague the classical smooth manifold $M$. These
unphysical infinities are on the one hand due to the fact that one
can pack an uncountable infinity of events into a finite
space-time volume, and on the other, due to the geometric
point-like nature of the latter. The manifold model, as an inert
classical geometric background stage on which fields propagate and
interact, must at least be revised in view of the pathological
nature of quantum gravity when treated as another quantum field
theory. Topoi and their relatives, {\em locales}, which are
pointless topological spaces modelling topological theories about
regions rather than points \cite{ref31}, are structures
well-suited not to significantly commit themselves to the
pathological geometric point-like character of a base space-time
manifold. Perhaps one could arrive at the `true' quantum topos of
Nature, on which a finite quantum theory of gravity can be
founded, by considering the pointless topos of the curved finitary
space-time sheaves of quantum causal sets instead of its classical
continuum relative ${\mathbf Shv}(M)$. In the same way that
${\mathbf Shv}(M)$ may be thought of as a universe of continuously
variable sets, so the topos of sheaves of quantum causal sets may
be thought of as a universe of dynamically variable quantum causal
sets---ones that vary due to a locally finite, causal and quantal
version of Lorentzian gravity.

Moreover, and in connection with our anticipation of the
development of a noncommutative topology and, as a result, of a
noncommutative sheaf theory adapted to it, the aforesaid quantum
topos of finitary sheaves of quantum causal sets is also expected
to provide a model of the missing structure in the following
analogy that has puzzled mathematicians for quite some time now
\cite{ref33}:

\begin{equation}\label{eq9}
\frac{\mathrm locales}{\mathrm quantales}=\frac{\mathrm
topoi}{\mathrm ?}
\end{equation}

\noindent For we have seen how the building of the Rota topology
on the quantum causal sets is akin to the definition of
noncommutative topological spaces {\it par excellence}, namely,
quantales \cite{ref9}, and there is a strong feeling among
mathematicians that the category of sheaves over a quantale is a
quantum version of the archetypal topos---the category of sheaves
over a locale \cite{ref27,ref34}.

Further to our noncommutative topological and sheaf-theoretic
epilogue, we mention that, since the incidence algebras modelling
quantum causal sets are graded non-abelian Polynomial Identity
(PI) rings, it would in principle be possible to develop a
noncommutative sheaf or {\em scheme} type of theory for the
non-abelian PI-ring (and differential module) localizations in
\cite{ref15,ref19,ref27}. Rigorous mathematical results, cast in a
general categorical setting, from the non-commutative algebraic
geometry of similar non-abelian schematic algebras and their
localizations \cite{ref35} are expected to deepen our physical
understanding of the dynamically variable non-commutative quantum
causal Rota topologies defined on the primitive spectra of quantum
causal sets \cite{ref27}.

However, such a possible noncommutative topological, scheme and
topos-theoretic application of quantum causal sets and their
sheaves to the problem of quantum gravity is still at its birth
\cite{ref36}.

We close the present paper with some remarks of Steve Selesnick
\cite{ref37} about the importance of developing a noncommutative
topological and sheaf-theoretic perspective on quantum space-time
structure and gravity:

\begin{quotation}

{\small ...One of the primary technical hurdles which must be
overcome by any theory that purports to account, on the basis of
microscopic quantum principles, for macroscopic effects (such as
the large-scale structure of what appears to us as space-time,
{\it ie}, gravity) is the handling of the transition from
`localness' to `globalness'. In the `classical' world this kind of
maneuver has been traditionally effected either
measure-theoretically---by evaluating largely mythical integrals,
for instance---or geometrically, through the use of sheaf theory,
which, surprisingly, has a close relation to topos theory. The
failure of integration methods in traditional approaches
to quantum gravity may be ascribed in large measure to the
inappropriateness of maintaining a manifold---a `classical'
object---as a model for space-time, while performing quantum
operations everywhere else. If we give up this classical manifold
and replace it by a quantal structure, then the already
considerable problem of mediating between local and global (or
micro and macro) is compounded with problems arising from the
appearance of subtle effects like quantum entanglement, and more
generally by the problems arising from the non-objective nature of
quantum `reality'. Although there is a rich and now highly developed
mathematical theory of `noncommutative geometry' (which has had
considerable success in application to traditional
quantum field theories), a concomitant noncommutative sheaf theory seems to
have been slow in coming...}

\end{quotation}

\vskip 0.1in

\noindent {\bf Acknowledgments.} I wish to thank Chris Isham, Jim Lambek,
Tasos Mallios, Chris Mulvey, Steve Selesnick, Rafael Sorkin, Freddy Van
Oystaeyen and Roman Zapatrin for numerous exchanges about quantum causal
sets, their sheaf-theoretic localizations, as well as their potential
noncommutative scheme and topos-theoretic generalizations. This
work was supported by an EU Marie Curie Postdoctoral Research
Fellowship held at Imperial College, London, United Kingdom.


\begin{thebibliography}{9999}
\bibitem{ref26} The `inner product problem' is the problem of fixing the inner product
in the Hilbert space of states by requiring that it is invariant
under ${\mathrm{Diff}}(M)$, while the `problem of time' is the
problem of requiring that the dynamics is encoded in the action of
${\mathrm{Diff}}(M)$ on the space of states.
\end{thebibliography}
\end{document}